\documentclass{bjour}
\usepackage{amsmath}

\newcommand{\ket}[1]{\ensuremath{|#1\rangle}}
\newcommand{\bra}[1]{\ensuremath{\langle#1|}}
\newcommand{\braket}[2]{\ensuremath{\langle#1|#2\rangle}}
\newcommand{\half}{\mbox{$\frac{1}{2}$}}
\newcommand{\tr}{\ensuremath{\text{tr}}}
\begin{document}
%
\authorrunninghead{Bowdrey and Jones}
\titlerunninghead{NMR Rotor Fidelity}
\title{A Simple and Convenient Measure of NMR Rotor Fidelity}
\author{M.~D. Bowdrey}
\affil{Oxford Centre for Quantum Computation, Clarendon
Laboratory, Parks Road, Oxford OX1~3PU, UK}
\email{mark.bowdrey@qubit.org}
\author{J.~A. Jones\thanks{To whom correspondence should be addressed at
the Clarendon Laboratory.}} \affil{Oxford Centre for Quantum
Computation, Clarendon Laboratory, Parks Road, Oxford OX1~3PU, UK,
and \\ Oxford Centre for Molecular Sciences, New Chemistry
Laboratory, South Parks Road, Oxford, OX1~3QT, UK}
\email{jonathan.jones@qubit.org}

\abstract{We describe a simple scheme for calculating the fidelity
of a composite pulse when considered as a universal rotor.}

\keywords{NMR, quantum computation, composite pulse, universal
rotor, fidelity}

\begin{article}
\section{Introduction}
Composite pulses \cite{Levitt:1979, Levitt:1986, Freeman:1997}
play an important role in many NMR experiments, as they allow the
effects of experimental imperfections, such as pulse length errors
and off-resonance effects, to be reduced.  In conventional NMR
experiments these pulses are used to implement particular
transformations on the Bloch sphere, such as inversion, and their
quality can be easily assessed by determining the efficiency with
which some known starting state is transferred to the desired
final state.  Furthermore the transfer efficiency can be both
calculated \emph{and} measured, allowing a simple comparison
between theory and experiment.

An alternative approach to composite pulses, developed by Tycko
\cite{Tycko:1983}, seeks to design \emph{general rotors}, that is
pulses which perform well for \emph{any} initial starting state.
Composite pulses of this kind, which are sometimes called Class A
composite pulses \cite{Levitt:1986}, are rarely (if ever) needed
for conventional NMR experiments, but are useful in NMR
implementations of quantum computation \cite{Cory:1996, Cory:1997,
Gershenfeld:1997, Jones:1998a, Jones:1998d}, where they act to
reduce systematic errors in quantum logic gates
\cite{Cummins:2000a}.  With pulses of this kind conventional
measures of transfer efficiency are inappropriate, and it is
necessary to consider the overall fidelity of the composite pulse
sequence when viewed as a general rotor.

A solution to this problem was provided by Levitt
\cite{Levitt:1986}, who defined the rotor fidelity by the dot
product of the quaternions describing the composite pulse and the
desired ideal rotation, and it is this approach which has been
used to date \cite{Cummins:2000a, Cummins:2001a}.  This
definition, however, has one major disadvantage: while it is
fairly easy to calculate it cannot be measured experimentally.  It
would be desirable to find a measure of rotor fidelity which, like
conventional quality measures, permits theory and experiment to be
compared.

One possible approach would be to use a conventional measure of
pulse sequence quality and to average this over a range of
starting states.  Intuitively this seems reasonable, but it is not
clear that such a method has any formal basis.  Here we show how
that this approach can, in fact, be derived from the definition of
propagator fidelity widely used in quantum information theory
\cite{Nielsen:2000}.

\section{Results}
A reasonable measure for the similarity of two pure quantum states
is the square of the overlap between them,
\begin{equation}
|\braket{\psi_1}{\psi_2}|^2.
\end{equation}
State overlap also provides a means to compare two different
propagators, $U$ and $V$, acting on the same state by considering
the overlap of their final states.  The \emph{fidelity} of the two
propagators (that is, the extent to which they are the same) can
then be obtained by averaging over all initial states:
\begin{equation}\begin{split}
f & = \overline{|\bra{\psi} U^\dag V \ket{\psi}|^2} \\
& = \overline{\tr(U\ket{\psi}\bra{\psi} U^\dag V
\ket{\psi}\bra{\psi} V^\dag)}.
\end{split}
\end{equation}

When considering a single spin-\half\ particle this expression can
be greatly simplified.  Any pure state of such a particle
corresponds to a point on the surface of the Bloch sphere, and the
corresponding density matrix $\rho(\theta,\phi)$ can be expanded
as a sum of the conventional $I_x$, $I_y$, and $I_z$ product
operators \cite{Sorensen:1983} and $I_0$ (half the identity
matrix, more normally written $\half E$).
\begin{equation}
\begin{split}
\rho(\theta,\phi) & = \half \mathbf{1} + \half \left(
\begin{array}{cc}
\cos{\theta} & \sin{\theta}e^{-i \phi} \\
\sin{\theta}e^{i \phi} & -\cos{\theta} \end{array} \right)
\\
& = I_0 + \sin{\theta} \cos{\phi} I_x + \sin{\theta}
\sin{\phi} I_y + \cos{\theta} I_z \\
& =  \sum_{j = 0,x,y,z}c_j(\theta,\phi) I_j.
\end{split}
\end{equation}
The propagator fidelity can be obtained by integrating over the
Bloch sphere.
\begin{equation}
f = \frac{1}{4\pi}\int_{0}^{2\pi}\!\!\!\int_{0}^{\pi}
f(\theta,\phi) \sin{\theta}\, \mathrm{d}\theta\, \mathrm{d}\phi
\end{equation}
where the fidelity at any point on the sphere, $f(\theta,\phi)$,
is given by
\begin{equation}
\begin{split}
f(\theta,\phi)& =  \tr\left(U \sum_j c_j(\theta,\phi) I_j
U^\dag V \sum_k c_k(\theta,\phi) I_k V^\dag\right)\\
& =  \sum_{j,k} c_j(\theta,\phi)c_k(\theta,\phi)\tr\left(U I_j
U^\dag V I_k V^\dag\right).
\end{split}
\end{equation}

When integrated over the Bloch sphere, the coefficients of terms
containing different product operators sum to zero, while the
diagonal terms survive:
\begin{equation}
\frac{1}{4\pi}\int_{0}^{2\pi}\!\!\!\int_{0}^{\pi} c_i(\theta,
\phi) c_j(\theta, \phi) \sin{\theta} \,\mathrm{d}\theta\,
\mathrm{d}\phi = \frac{\delta_{ij}}{3} + \frac{2 \delta_{i0}
\delta_{j0}}{3}.
\end{equation}
Note that $I_0$ (which is a multiple of the identity matrix)
commutes with any propagator, and so
\begin{equation}
\begin{split}
\tr\left(U I_0 U^\dag V I_0 V^\dag\right) &=\tr\left(I_0 UU^\dag
I_0VV^\dag\right)\\
&=\tr\left(I_0^2\right)=\half.
\end{split}
\end{equation}
Hence the propagator fidelity is reduced to
\begin{equation}
\label{eq:fidelity} f = \frac{1}{2} + \frac{1}{3} \sum_{j = x,y,z}
\tr\left(U I_j U^\dag V I_j V^\dag\right).
\end{equation}

For an isolated spin-\half\ particle any unitary propagator can be
considered as a rotation, and the rotor fidelity of a pulse
sequence can be taken as the fidelity of the corresponding
propagator.  As NMR experiments are usually conducted at high
temperature, the spin is not described by a pure state, as assumed
above, but by a highly mixed state.  This state can, however, be
considered as a mixture of the unit matrix and a deviation density
matrix corresponding to a pure state,
\begin{equation}
\ket{\psi'}\bra{\psi'}=(1-\epsilon)\half\mathbf{1} + \epsilon
\ket{\psi}\bra{\psi}
\end{equation}
and in most situations we are only interested in the deviation
matrix $\ket{\psi}\bra{\psi}$.  (In the language of NMR quantum
computation this is referred to as the \emph{pseudo pure state}
approach \cite{Cory:1997,Jones:1998d}; this is also the approach
adopted in conventional NMR studies except that the contribution
of $I_0$ to the deviation matrix is rarely considered).

Examining the behaviour of the deviation matrix leads to the same
results as for a pure state except for a scaling factor of
$\epsilon^2$, and since $\epsilon$ is in effect a measure of
signal strength we follow common NMR practice and set
$\epsilon=1$.  Thus the rotor fidelity is given by the previous
expression for the propagator fidelity,
equation~\ref{eq:fidelity}. Finally we note that $\tr(U I_j U^\dag
V I_j V^\dag)$ is simply half of the efficiency with which the
propagator $V$ transfers the initial state $I_j$ to the desired
final state $U I_j U^\dag$. Thus, neglecting a scaling factor of
one half and an offset of one half, the rotor fidelity is equal to
the conventional transfer efficiency averaged over the three
starting states $I_x$, $I_y$ and $I_z$.

\section{Conclusions}
The propagator fidelity provides a simple and convenient measure
of rotor fidelity which can be used to assess the quality of NMR
composite pulses designed to act as general rotors.  Unlike the
quaternion measure introduced by Levitt \cite{Levitt:1986} the
propagator fidelity can be both calculated (using
equation~\ref{eq:fidelity}) and measured experimentally (by
averaging the conventional transfer efficiency over the three
starting states $I_x$, $I_y$ and $I_z$).

\begin{acknowledgment}

We thank E.~Galv\~{a}o, L.~Hardy, D.~Oi and T.~Short for helpful
conversations. M.D.B. thanks EPSRC (UK) for a research fellowship.
J.A.J. is a Royal Society University Research Fellow. This is a
contribution from the Oxford Centre for Molecular Sciences, which
is supported by the UK EPSRC, BBSRC, and MRC.

\end{acknowledgment}

\end{article}

\end{document}